# Additive manufacturing and performance of bioceramic scaffolds with different hollow strut geometries

## Abstract

Additively manufactured hollow-strut bioceramic scaffolds present a promising strategy towards enhanced performance in patient-tailored bone tissue engineering. The channels in such scaffolds offer pathways for nutrient and cell transport and facilitate effective osseointegration and vascularization. In this study, we report an approach for the slurry based additive manufacturing of modified diopside bioceramics that enables the production of hollow-strut scaffolds with diverse cross-sectional forms, distinguished by different configurations of channel and strut geometries. The prepared scaffolds exhibit levels of porosity and mechanical strength that are well suited for osteoporotic bone repair. Mechanical characterization in orthogonal orientations revealed that a square outer cross-section for hollow struts in woodpile scaffolds gives rise to levels of compressive strength that are higher than those of conventional solid cylindrical strut scaffolds despite a significantly lower density. Finite element analysis confirms that this improved strength arises from lower stress concentration in such geometries. It was shown that hollow struts in bioceramic scaffolds dramatically increase cell attachment and proliferation, potentially promoting new bone tissue formation within the scaffold channel. This work provides an easily controlled method for the extrusion-based 3D printing of hollow strut scaffolds. We show here how the production of hollow struts with controllable geometry can serve to enhance both the functional and mechanical performance of porous structures, with particular relevance for bone tissue engineering scaffolds.

# 1. Introduction

Large bone defects as a result of congenital malformations, tumors, and infections continue to pose serious clinical challenges. One promising approach towards the treatment of such defects is the implantation of bio-resorbable ceramic scaffolds into defects to promote and guide the formation of new bone tissue [1, 2]. 3D printing is a low-cost, rapid on-demand approach that enables the fabrication of structurally complex patient-specific scaffolds [3, 4]. Generally, 3D printed ceramic scaffolds have high macroscopic porosity in the form of large voids. However, most such scaffolds consist of solid struts that do not present a channel structure at finer length scales, which would not be conducive to rapid vascularization and inward growth of host tissue [5, 6]. This is particularly problematic for large bone defects, as mineralization occurs primarily in the external region of the scaffold, while tissue formation is inhibited in the internal region of the implant due to inadequate transport pathways [7]. The fabrication of scaffolds with a channel structure based on hollow struts is a promising strategy to improve nutrient and cell transport in regenerating tissues and to accelerate vascularization and tissue growth deep within the scaffold [8]. The interplay between the bioactivity of ceramic scaffold materials and the transport pathways afforded by the hollow channels has been recognized as a key aspect in governing the overall remedial performance of these promising scaffolds [9, 10]. The development of practical methods towards the realization of high-performance hollow-strut scaffolds is a valuable objective and a key motivator of the present work.

Several studies have shown that the pore geometry and dimensions can significantly influence tissue regeneration processes in scaffolds; in particular, pore sizes higher than 300 μm have been shown to allow vascularization and new bone formation [11-13]. However, most studies into hollow-strut scaffolds involved only tubular structures with hollow cylindrical members. A logical approach that merits exploration is the fabrication of hollow-strut scaffolds with varying outer (strut) and inner (channel) geometries. A systematic study of different strut geometries and their fabrication is presented here to inform and guide the improved design and fabrication of such scaffolds.

In the past few years, several studies have presented methods to form hollow channels in printed scaffolds [14-17]. In several methods, hollow channels in scaffolds were achieved using sacrificial core materials that were subsequently removed [16]. Such methods are somewhat limited with regard to achievable structure and materials. Scaffolds with hollow channel structures have recently been manufactured using coaxial nozzle-assisted 3D printing. Most such methods rely on a coaxial nozzle with two nested needles (slurry in the outer port, and air in the inner one) to create a continuous hollow channel strut, generally confined to cylindrical geometries [17]. It can be seen from previous studies that the method of constructing coaxial nozzles by inserting a second needle with a smaller diameter into a larger needle cannot guarantee a well-centered hollow channel and is prone to deformation. There is a clear motivation to identify more efficient and controllable methods of 3D printing hollow scaffolds.

In this work, we reveal an easily controlled and cost-effective method to fabricate scaffolds consisting of hollow struts with different external and internal geometries. To this end, we designed and produced a new type of integrated extrusion nozzles, which were further applied within a robocasting approach, thus allowing the direct extrusion of hollow struts with variable geometries and cross-sections. Scaffolds were based on diopside, with 1 at.% substitution of magnesium by copper (Cu-DIO, $CaMg_{0.99}Cu_{0.01}Si_2O_6$), as this bioceramic was recently shown to exhibit an exceptional combination of both mechanical properties and bioactivity, in terms of fracture toughness, cell proliferation, *in vitro* angiogenesis and antibacterial activity [18]. The mechanical properties of hollow-strut scaffolds are studied here experimentally and simulated by finite element analysis (FEA). The potential application of hollow struts for bone repair is investigated, as well. The method we present here for fabricating hollow-strut scaffolds is inexpensive and easily controllable, showing great utility in extrusion-based additive manufacturing of complex structures with improved mechanical properties.

## 2. Materials and methods
### 2.1. Synthesis of Cu-DIO powders

The first step towards forming a printable slurry is the synthesis of ceramic powder. Here, copper doped diopside powder with a stoichiometry of $CaMg_{0.99}Cu_{0.01}Si_2O_6$ was synthesized by a co-precipitation method. Briefly, magnesium chloride hexahydrate ($MgCl_2 \cdot 6H_2O$, >98%, Carl-Roth), calcium chloride ($CaCl_2$, >98%, Carl-Roth), and copper chloride ($CuCl_2$, 99%, Sigma-Aldrich) were added to ethanol. After complete dissolution, tetraethyl orthosilicate (($C_2H_5O)_4Si$, TEOS, VWR Chemicals) was added in a molar ratio of 1 : 0.99 : 0.01 : 2 (Ca : Mg : Cu : Si) and stirred for 2 h. For precipitation, 25% ammonia solution (Carl-Roth) was added and stirred overnight. The co-precipitated product was then centrifuged and washed before being dried overnight at 80°C and calcined in air for 2 h. Finally, the synthesized Cu-DIO powder was collected and sieved to a particle size smaller than 125 μm for subsequent use.

### 2.2. Design and production of extrusion nozzles

To produce struts with desired dimensions and different cross-sections, customized nozzles with various outlet configurations were required. Fig. 1 shows schematically the design of nozzles. The modelling of nozzles was performed using Pro/Engineering software. To solve the prevalent problem of uncoaxiality in nozzles with needle nested structure, nozzles were fabricated using stereolithography (SLA) by means of an SLA 3D printer (Form 2, Formlabs) with clear photoreactive liquid resin (Clear Resin V4, Formlabs). The stereolithographic fabrication of these nozzles was done with a layer height of 50 μm. As shown in Fig. 1b and Fig. 1c, the extruding nozzles were designed with a female Luer inlet for connecting with the cartridge. At the end of the outlet, a mandrel with a length of 10 mm was fixed with two small spatially orthogonal bridges. Notably, the mandrel protrudes 0.3 mm out of the outlet, in order to achieve a better modeling effect. Naturally, the outlet geometry of these nozzles governs the external cross-section of printed struts, while the inner mandrel geometry determines the geometry of the hollow longitudinal channel within the struts. The basic nozzle design consisted of a round outlet (1.70 mm in diameter) and a round mandrel (0.65 mm in diameter). All the nozzles were designed to have the same cross-sectional area. The nozzles can be integrally formed during the SLA printing, resulting in good coaxiality between the mandrel and the outlet. A total of 10 nozzles were fabricated and their cross-sectional views are shown in Fig. 1d and Fig. 1e, followed by rinsing with isopropanol and post-processing UV curing. After printing and post-treatment, the various nozzles were obtained with satisfactory resolution and fidelity (Fig. 1f). Nozzles with round, triangle, square, pentagonal, and hexagonal mandrels were designated as In-r, In-t, In-s, In-p, and In-h, respectively. Similarly, nozzles with different outlet geometries were named as Out-t, Out-s, Out-p, and Out-h, respectively. A basic tubular extrusion nozzle without a mandrel was labeled as Solid. The bioceramic scaffolds printed by corresponding nozzles were named in the same way. 3D models of the fabricated nozzles are provided as supplementary information here, as .stl files.

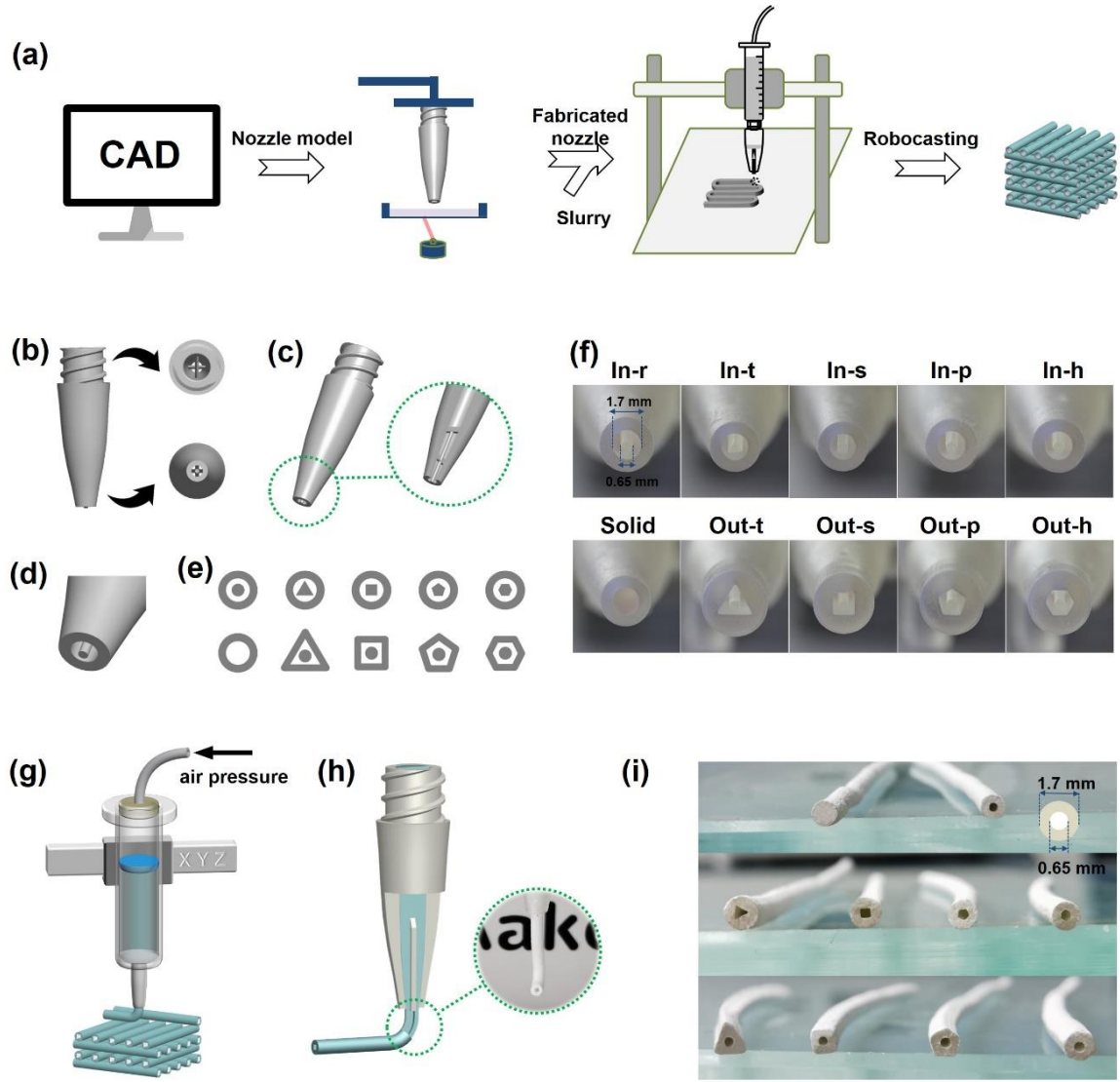

Fig. 1. (a) Schematic illustration of the manufacturing approach towards hollow-strut scaffold developed and applied in this study; (b-d) The structure of extruding nozzles; (e) The cross-sections of the extruding nozzles; (f) Tips of fabricated extruding nozzles under magnification; (g-h) Schematic illustration of scaffold fabrication and hollow strut formation; (i) Extruded hollow struts (green body) with different cross-sectional shapes.

## 2.3. Preparation of slurry and robocasting of scaffolds

In this study, robocasting was used to fabricate woodpile-type scaffolds, which is regarded as a representative and readily reproducible scaffold geometry. To prepare a suitable printable slurry with shear thinning behavior, 10.0 g Cu-DIO powder, 15.0 g 20 wt% Pluronic F127 (Sigma-Aldrich), 0.3 g 1-octanol (Merck), and 0.5 g sodium alginate (low viscosity, Alfa Aesar) were mixed and homogenized. The prepared slurry was transferred into a syringe (Vieweg GmbH, German), to which our customized nozzles were coupled. The syringe was tapped vigorously to remove bubbles and fixed onto a robotic deposition 3D printer, which was modified and re-equipped with an air-pressure controller based on the Ultimaker 2+ printer (Ultimaker BV, Netherlands) (Fig. 1g). As there was a fixed core in the center of the outlet, the slurry can only be extruded from the space between the outlet and the mandrel, and therefore forming a hollow-strut filament (Fig. 1h). It can be seen in Fig. 1i that ten filaments with different cross-section shapes were successfully extruded in accordance with the designed nozzles, allowing the further fabrication of scaffolds with different hollow struts. The green bodies were then transported to a covered container with a small opening for slow drying for 4 days, during which time they were flipped over at least once. This is crucial to prevent cracking and ensure homogeneous drying. Finally, the dried green bodies were sintered for 3 h at 1250°C. The final scaffold is obtained by cutting off the ends of the sintered scaffold using a horizontal precision diamond wire saw (WELL Diamond Wire Saw 3400) to expose the cross-section.

## 2.4. Rheological characterization

The properties of slurries play a central role in the printability of materials and structures in extrusion-based additive manufacturing methods, which is particularly true for the extrusion of hollow geometries. An advantage of using a fixed mandrel extrusion nozzle in the present work was found to be improved channel-strut co-axiality and the ability to utilize more viscous slurries in comparison with nested-needle setups. In order to study the suitability of the slurries used here for extrusion-based processing, rheological tests were carried out on a rotational rheometer (AntonPaar, Physics MCR 301, Austria) equipped with a 2° steel cone plate (25 mm diameter) at 25°C. 40 wt% Pluronic F127 aqueous solution was used as control. Samples were equilibrated for 8 min before running. A shear rate sweep was conducted from $10^{-3}$ to $5 \times 10^2$ s$^{-1}$. An oscillatory strain sweep was performed to record the storage modulus (G′) and loss modulus (G″) of the materials in the range of 0.01% to 200% at 1 Hz. To simulate the robocasting process, a shear recovery test with three switching steps of low (0.01%)-high (200%)-low (0.01%) shear strain was studied. Based on the testing result of three switching steps, the loss factor tan δ (tan δ = G″/G′) was calculated for all samples. Besides, a shear stress sweep was also performed from 1 Pa to $3 \times 10^3$ Pa to investigate the yielding behavior of the slurry.

## 2.5. Characterization of scaffolds

X-ray diffraction (XRD) patterns of sintered Cu-DIO scaffolds were collected by an X-ray diffractometer (D8 advance, Bruker AXS, USA) at 35 kV and 40 mA using CuKα radiation in 3 s/0.02° steps. Energy dispersive X-ray (EDX) elemental mapping was performed on the cross-section of a representative In-r scaffold using an EDX spectrometer (Thermo Fisher Scientific Inc., USA). The scaffold's cross-section and the channel were exposed by cutting off the ends using a diamond wire saw. The images of the scaffolds were captured by a digital camera (EOS 700D, Canon, Japan) equipped with a macro lens (EF-S 60mm f/2.8 Macro USM, Canon, Japan). Scanning electron microscopy (SEM, LEO Gemini 1530, Zeiss, Germany) was used to observe a representative cross-section of a single strut in In-r scaffold. Optical microscopy (DMC2900, Leica, Germany) was used to examine the sintered scaffolds' hollow channel morphologies. An LS 13320 laser diffraction particle size analyzer (Beckman Coulter, USA) was used to analyze the particle size in an ultrasonically dispersed aqueous solution. The scaffold shrinkage was calculated from the width, length, and height of the scaffold before and after sintering. The weight and volume of the scaffolds were used to calculate their apparent density. The open porosity of the scaffolds was measured using the Archimedes method with the following equation (1):

$$\text{Porosity (\%)} = \frac{W_w - W_d}{W_w - W_s} \times 100 \quad (1)$$

where $W_w$ represents the weight of the scaffold with water, $W_d$ represents the dry weight of the scaffold, and $W_s$ represents the wet weight of the scaffolds suspended in water. Six replicate experiments were performed.

## 2.6. Mechanical testing

The compressive strength of Cu-DIO scaffolds ($8 \times 8 \times 12$ mm$^3$) comprised of different strut types was determined by compressing the scaffold at a crosshead speed of 0.5 mm/min using a mechanical testing machine (Zwick/Roell, Germany) equipped with 20 kN load cell. The compression load was applied in two directions: transverse and parallel to the struts' direction of travel. The stress-strain curves were recorded throughout the testing. The measurement was conducted in twelve replicates. The ratio of compressive strength to apparent density was used to calculate the specific strength of the scaffold. Four-point bend testing was conducted on sintered scaffolds with dimensions of $45 \times 5 \times 4$ mm$^3$ at a crosshead speed of 0.5 mm/min. The measurement was conducted in twelve replicates. During the testing, the load-displacement curves were recorded, and the four-point bend strengths (σ) were determined using the following equation [19] (2):

$$\sigma = \frac{3F(L - L_1)}{2bd^2} \quad (2)$$

where F is the force, L is the support rollers' span, $L_1$ is the loading span separation, d is sample thickness, and b is sample width.

## 2.7. Finite Element Analysis (FEA)

ANSYS workbench (ANSYS, Inc., Canonsburg, PA, USA) was used to simulate stress distribution in the scaffolds under compression and bending. Material properties of diopside were assigned to the modelled parts. A Poisson's ratio of 0.35 and Young's modulus of $1.7 \times 10^{11}$ Pa were used for the material according to a previous report [20]. To simulate compression, a unit cell of each scaffold geometry was considered to have flat upper and lower surfaces. The models were designed and assembled in Pro/Engineering, and then imported into ANSYS workbench. Rigid parallel plates were simulated on the upper and lower surfaces of the scaffold unit to facilitate uniaxial compressive load conditions. The upper rigid plate was subjected to a 300 N compressive force, while the bottom rigid plate was fixed [21, 22]. For the four-point bending simulation, the scaffolds with sizes of $45 \times 5 \times 4$ mm$^3$ based on the four-point bending experiment were modelled and imported similarly. And four semicircular rods with 3 mm diameter were also created, the upper two were applied with 50 N downward load while the bottom two were fixed. The contact between the scaffolds and the rods was assumed to be frictionless. All the samples had identical boundary conditions. Mesh sensitivity analysis was carried out by incrementally decreasing mesh size until a mesh size-independent result was reached. Subsequently, the scaffolds were meshed with an element size of 0.2 mm, while this value for rigid plates and semicircular rods was 0.5 mm. Equivalent stresses (Von Mises stress) or total deformations were analyzed for the specimens.

## 2.8. *In vitro* cell experiment

The impacts of Cu-DIO on cell behavior were studied using human umbilical vein endothelial cells (HUVECs, ATCC CRL-1730), and human osteogenic sarcoma cells (Saos-2 cells, Department of Periodontology, Oral Medicine and Oral Surgery, Charité), which were cultured in a humidified incubator (5% $CO_2$, 37ºC). For cell culture, complete medium was used and replaced every other day. Specifically, MCDB 131 medium (Gibco, Thermo Fisher Scientific) with 2 mM L-Glutamine and 10% fetal bovine serum (FBS) was used for HUVECs; Dulbecco's Modified Eagle medium (DMEM, High Glucose, Biowest) with 2 mM L-Glutamine and 15% FBS was used for Saos-2 cells. Cu-DIO scaffold extract was obtained by soaking 2 g of scaffold in 10 mL serum-free medium at 37°C for 24 h according to ISO 10993-12 and 10993-5, the supernatant was then collected and sterilized using a 0.22 μm filter (Millipore), subsequently, 10% FBS was supplemented.

**Cell viability assay**

2,3-bis (2-methoxy-4-nitro-5-sulfophenyl)-2H-tetrazolium-5-carboxanilide sodium salt (XTT sodium salt, Alfa Aesar) was used to assess cell viability following exposure to Cu-DIO scaffold extract, as described by Wu et al. [23]. Briefly, after seeding the cells in 96-well culture plates, the supernatant in the experimental well was changed for 100 μL scaffold extract, and XTT solution with phenazine methosulfate (PMS, AppliChem, Darmstadt, Germany) (finally 0.3 mg/mL and 2.5 μg/mL, respectively) were added to each experimental well after 1, 3, and 7 days of incubation, respectively. After 4 h of incubation, the absorbance of the supernatant at 450 nm was determined using a microplate reader (Tecan Austria GmbH). As a background, fresh medium was tested under the same conditions. Six copies of the test were carried out.

To assess cell viability on scaffolds of different shapes, HUVECs and Saos-2 cells were stained using a live/dead staining kit (L3224, Thermo Fisher Scientific). Briefly, after seeding cells on the scaffold for 48 h, the scaffolds were rinsed with phosphate buffered saline (PBS), followed by the addition of 2 μM ethidium homodimer-1 and 2 μM calcein AM to the experimental wells and incubation in an incubator for 15 min. Fluorescence microscopy (Observer Z1, Zeiss) was then used to observe fluorescence in three repetitions. Green fluorescence represents living cells, while red fluorescence indicates dead cells.

In order to evaluate the cell survival inside the hollow struts, Saos-2 cells were seeded onto the outer surfaces and the inside channels of the hollow struts of In-r scaffolds. After 7 days of culture, the scaffolds were cut using a diamond wire saw to reveal the cross-sections of the channels. Then the live/dead assay and fluorescence microscopy were performed.

**Western blot analysis**

The expressions of angiogenesis-related proteins were assessed using western blot analysis. HUVECs were harvested with lysis buffer after incubation with Cu-DIO scaffold extract and pure medium for 3 days, respectively. Protein concentrations were measured using a bicinchoninic acid (BCA) reagent. The total protein (50 μg) was separated using the SDS-polyacrylamide gel electrophoresis gel (SDS-PAGE, BIO-RAD, United States) and transferred to a polyvinylidene fluoride (PVDF) membrane. The membrane was incubated overnight at 4ºC with platelet endothelial cell adhesion molecule-1 (CD31) antibody (1 : 1000, Cell Signaling, Massachusetts, USA) and further the relevant secondary antibody (goat anti-mouse, 1 : 100000, Abcam, Cambridge, UK). The immunoblotting was also performed for α-Actin serving as an internal loading control. The protein bands were visualized using the enhanced chemiluminescence substrate kit (Pierce ECL western blotting substrate, Thermo Scientific, USA) and captured with a BIO-RAD imaging system (ChemiDoc MP, United States). The result was further analyzed using Image Lab software (BIO-RAD).

## 2.9. Production of polymer-infiltrated scaffolds

The effective production of hollow-strut ceramic scaffolds with uninterrupted inner channels further represents a promising new avenue towards the production of core-shell members in scaffolds. To demonstrate this approach to processing core-shell polymer/ceramic composites, the channel of a representative In-r scaffold was filled with polycaprolactone (PCL, Mn 80000, Sigma-Aldrich). PCL was chosen because of its biodegradability, biocompatibility, and mechanical strength, which make it popular in bone tissue engineering [24]. To achieve this infiltration, PCL was dissolved in a toluene solution at a concentration of 20 vol% at 60°C. The PCL solution was then manually injected into each strut until it was entirely filled using a 30G (Vieweg GmbH) fine needle, and then left at room temperature for 24 h to ensure adequate evaporation of the toluene solvent. The morphology of the PCL-filled In-r scaffold was observed by optical microscopy and the compression tests were conducted on the scaffold.

## 2.10. Statistical analysis

In this work, all data were expressed as a mean value with a standard deviation (SD). GraphPad Prism 8 software was used to carry out the statistical analysis. The differences between experimental groups were analyzed, where *$p$ < 0.05 and **$p$ < 0.01 denote that the results are significantly different.

# 3. Results
## 3.1. Slurry rheology

The properties of the slurry must be considered while producing hollow scaffolds using extrusion-based 3D printing techniques, as it fundamentally affects the printing results [25, 26]. To investigate the printability of the Cu-DIO slurry, rheological analyses were performed with the results given in Fig. 2. In rheological characterization, a 40 wt% Pluronic F127 aqueous solution was introduced as a control since it is widely utilized in extrusion-based 3D printing and demonstrated to have excellent printability [27, 28]. Within the testing range of shear rate, the Cu-DIO slurry showed a moderately higher viscosity than the Pluronic F127 (Fig. 2a). And the viscosity of both decreased rapidly with an increasing shear rate. It is a typical phenomenon of pseudoplastic fluids, indicating the shear-thinning behavior of the slurry, which is essential for robocasting [29, 30]. This is consistent with the result of the dynamic strain sweep test; two materials showed a very similar trend under a gradually increased shear strain (Fig. 2b). They showed a solid-like behavior under small strain as elastic modulus (G') was much higher than loss modulus (G") when the shear strain was below 1%. And they underwent a transition at around 5% shear strain from solid-like to liquid-like behavior, where G' became smaller than G". And therefore, the slurry and Pluronic F127 can be extruded smoothly from the nozzle under shearing.

A low-high-low shear strain 3-step switching test was applied to simulate the robocasting procedure. It is found that both Cu-DIO slurry and Pluronic F127 rapidly transitioned into the liquid-like state (G' dropped dramatically by around 3 and 2 order of magnitude for the slurry and Pluronic F127) when the shear strain increased from 0.01% to 200%, which represented the slurry in the cartridge and under extrusion, respectively (Fig. 2c). Then a small shear strain of 0.01% was set to mimic the deposition stage. As a result, it is shown that both materials can recover into a solid-like state instantly under a small strain as the G' significantly exceeded G". Correspondingly, the loss factor tan δ, which is the ratio of G″ to G′, is used to describe viscoelastic behavior. tan δ values of both slurry and Pluronic F127 increased almost instantly under shearing and dropped immediately after shearing (Fig.

2c). This is a characteristic transition of a solid-like (tan δ < 1) to and from a liquid-like (tan δ > 1) state. When the shearing stopped, the loss factor of both materials very rapidly returned to almost the same level as before shearing. Notably, it is indicated that recovery was more rapid in Cu-DIO slurry than in Pluronic F127. In addition, a shear stress ramp test was conducted, with results shown in Fig. 2d. Compared to unprintable materials, the slurry in this work and the Pluronic F127 showed a gradual increase in shear strain upon increasing shear stress below 500 Pa [31]. Notably, the slurry was measured to have a lower shear strain compared to Pluronic F127 under the same shear stress. On the basis of these rheological findigs, the slurry we used in the fabrication of the scaffolds reported here is considered to be highly suitable for extrusion based printing. The rapid recovery of a solid-like state post-deposition is particularly important in order to maintain the deposited hollow channel structures without collapsing.

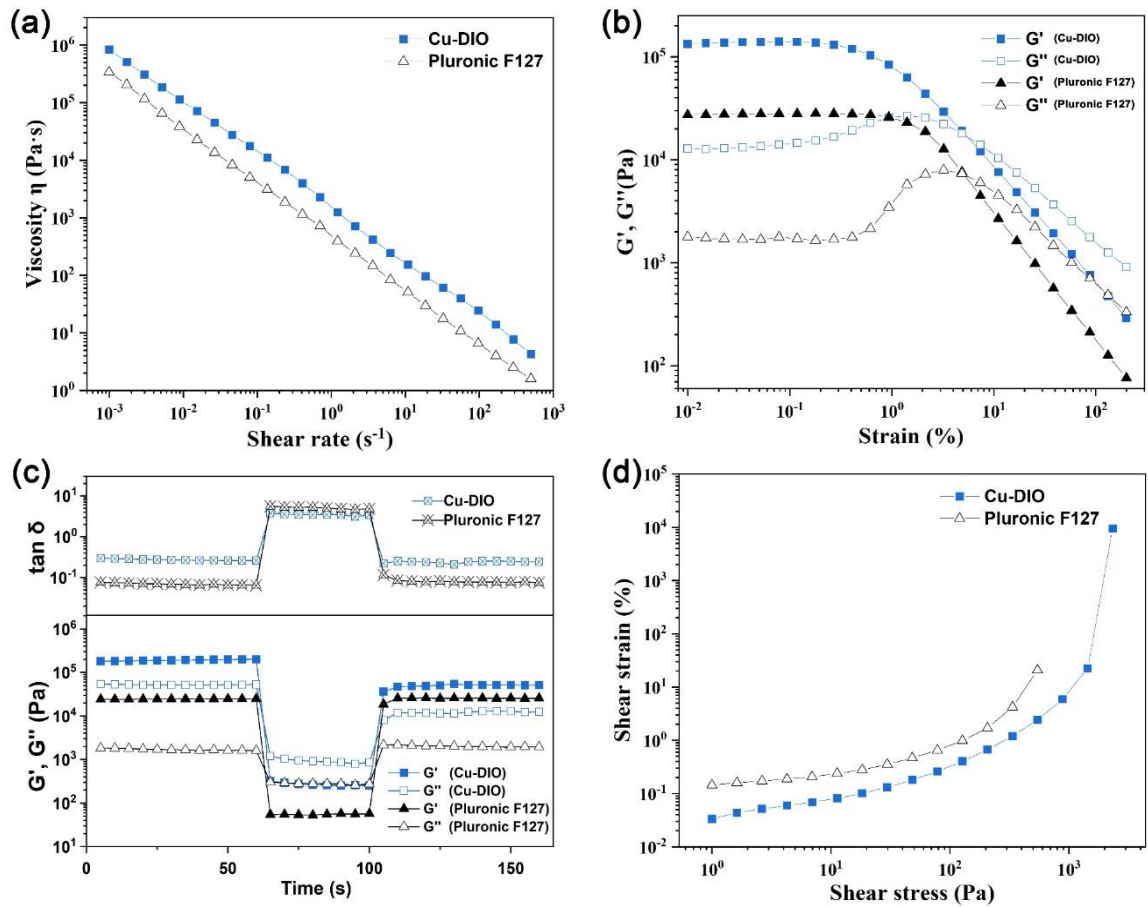

Fig. 2. Rheological characterization of Cu-DIO slurry and 40 wt% Pluronic F127 aqueous solution. Different test modes were used: (a) shear rate sweep, (b) shear strain sweep, (c) 3 steps shear strain switching test, and (d) shear stress sweep.

## 3.2. Structural and elemental composition of scaffolds

An XRD pattern of the sintered scaffold is presented in Fig. 3a. As expected, the pattern is consistent with the presence of single-phase diopside (PDF#72-1497), indicating that the binders were burned off and the incorporation of copper ions into diopside at this concentration did not lead to the formation of secondary crystalline phases. As shown in Fig. 3b, EDX was utilized to examine the elemental composition and distribution of the scaffold. The elements Si, O, Mg, Ca, and Cu were found to be evenly distributed over the hollow scaffold's cross-sectional surface.

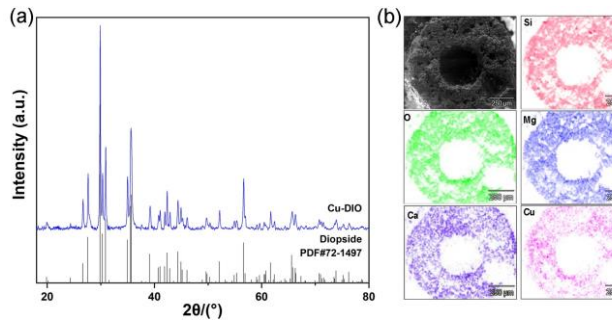

Fig. 3. Structural characterization of Cu-DIO scaffold after sintering at 1250°C. (a) XRD pattern; (b) the representative cross-sectional EDS elemental mapping images of In-r scaffold.

Scaffolds with various hollow strut geometries can be directly extruded by robocasting using the designed nozzles. After sintering, the scaffold with hollow channels had a significantly higher porosity of approximately 84% compared to the solid scaffold (73%), corresponding to apparent densities of $1.145 \pm 0.052$ g/cm$^3$ and $1.369 \pm 0.045$ g/cm$^3$, respectively. Despite shrinkage during sintering, scaffolds still maintained their designed geometries. Specifically, the shrinkage was $35.58 \pm 1.39\%$ and $35.70 \pm 1.51\%$ in the X and Y directions, respectively, while the shrinkage in the Z direction was $44.48 \pm 1.59\%$. Fig. 4a displays the images of an *in silico* and a sintered In-r scaffold after cutting so as to see the interior cross-sections of the hollow struts. A representative cross-sectional SEM image of In-r after sintering given in Fig. 4b and 4c shows that the round inner channels were well preserved (inner diameter ≈ 430 μm, outer diameter ≈ 1.1 mm), and internally connected. The high magnification image (Fig. 4d) shows that the sintered Cu-DIO materials here had a typical densified polycrystalline microstructure. The median particle size measured by particle size analysis was 8.1 μm (Fig. S1), which is higher than that shown in Fig. 4d, most likely due to particle agglomeration during the measurement. For visualization, the cross-section of a representative In-r scaffold was cut to show the channel. Fig. 4e shows the open hollow channel within the struts and the upper and lower layers were tightly bonded together, without obvious interlayer defects. The optical micrographs in Fig. 4f-o show the cross-sectional morphology of the woodpile structure scaffold with precisely defined shapes: Solid, In-r, In-t, In-s, In-p, In-h, Out-t, Out-s, Out-p, and Out-h. It is demonstrated from Fig. 4f-o that the geometry of the cross-section of each strut matches well with the designed nozzle model, with a high concentration of internal channels and no significant structural deformation.

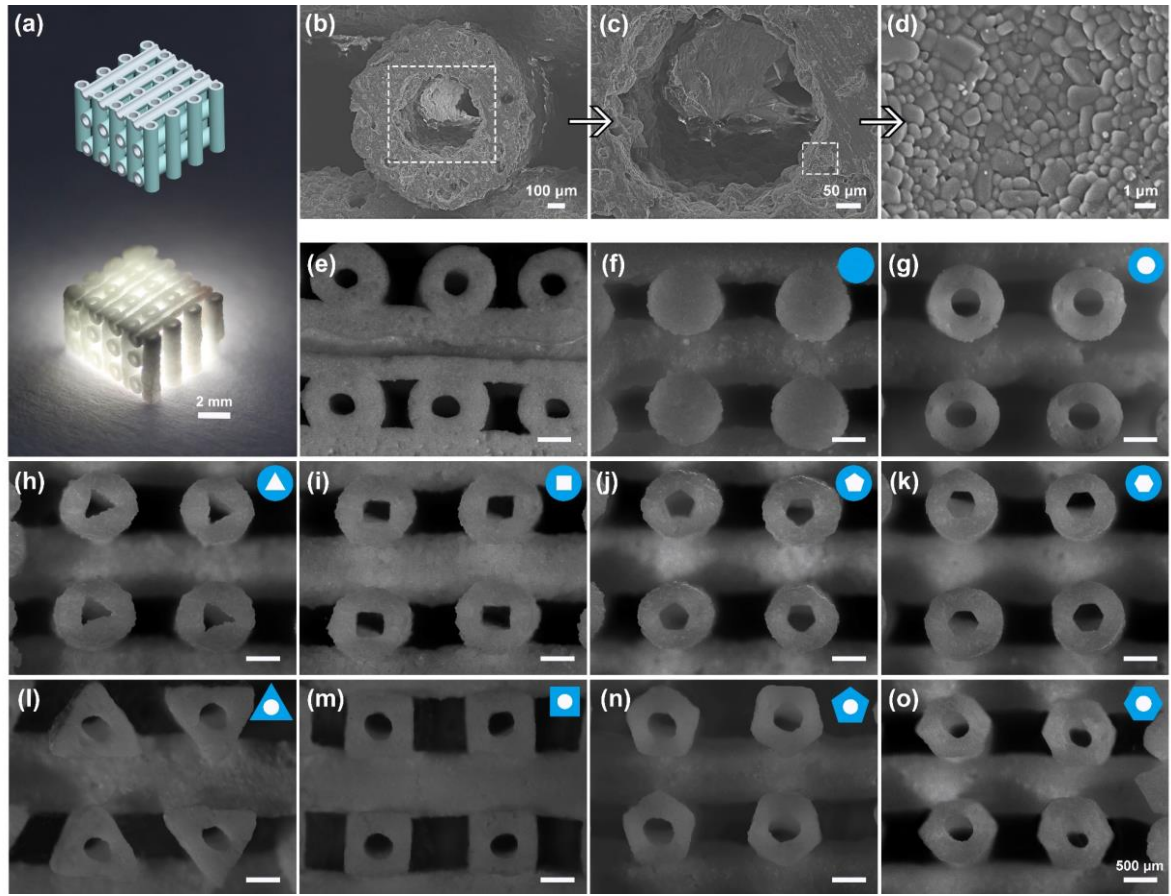

Fig. 4. Structural characterization. (a) An overall image of representative In-r after cutting to show the interior cross-sections of the hollow struts; (b-d) Representative cross-sectional SEM image of In-r at different magnifications after sintering at 1250°C; (e) After cutting, a representative In-r scaffold shows the internal channels; Representative cross-sectional optical image of scaffolds: (f) Solid, (g) In-r, (h) In-t, (i) In-s, (j) In-p, (k) In-h, (l) Out-t, (m) Out-s, (n) Out-p and (o) Out-h.

### 3.3. Mechanical performance of scaffolds

To examine the relationship between strut geometry and mechanical performance, compression and four-point bending tests were carried out on sintered Cu-DIO scaffolds. As shown in Fig. 5a and b, compressive strength was tested in the transverse direction and in parallel to the struts. Unsurprisingly, relative to the denser solid scaffold, the compressive strength in both directions was somewhat lower for almost all geometries of hollow-strut scaffolds. Interestingly, Out-s, corresponding to a square external cross-section with a circular channel, showed the highest compressive strength in the transverse direction, which was 1.2 times higher than that of the solid scaffold, despite having a lower density. In terms of specific strength, in other words, the strength-to-weight ratio of the scaffold structures studied here, it can be seen that hollow strut printing with square member geometry (Out-s) enhances performance with levels of specific compressive strength in the transverse direction of approximately 34% higher than solid round struts, as shown in Table S1. When tested in the parallel direction, hollow-strut scaffolds are weaker than the solid scaffold, and no statistically significant differences in compressive strength among the hollow strut geometries, with the Out-s scaffold having only a slightly higher compressive strength. In the parallel direction, solid strut scaffolds have slightly higher specific strength compared to Out-s, 38.01 ± 4.28 vs. 36.93 ± 4.70 MPa·cm$^3$/g, respectively.

For hollow-strut scaffolds with non-square outer geometry, mechanical performance was generally poorer than that of the solid-strut structure. In both orientations, In-t scaffolds, having a triangular channel in a circular strut, showed the lowest compressive strengths. Fig. S2a and Fig. S2b show representative stress-strain curves of the scaffold under compression in two directions, respectively. Under transverse compression, scaffolds showed a typical progressive failure characterized by a fluctuating load up to peak strength (Fig. S2a). When scaffolds were compressed parallel to the direction of the struts, a linear elastic compression was observed for both hollow and dense scaffolds, followed by a sharp decrease in load associated with catastrophic fracture, most likely by shearing or buckling of members (Fig. S2b).

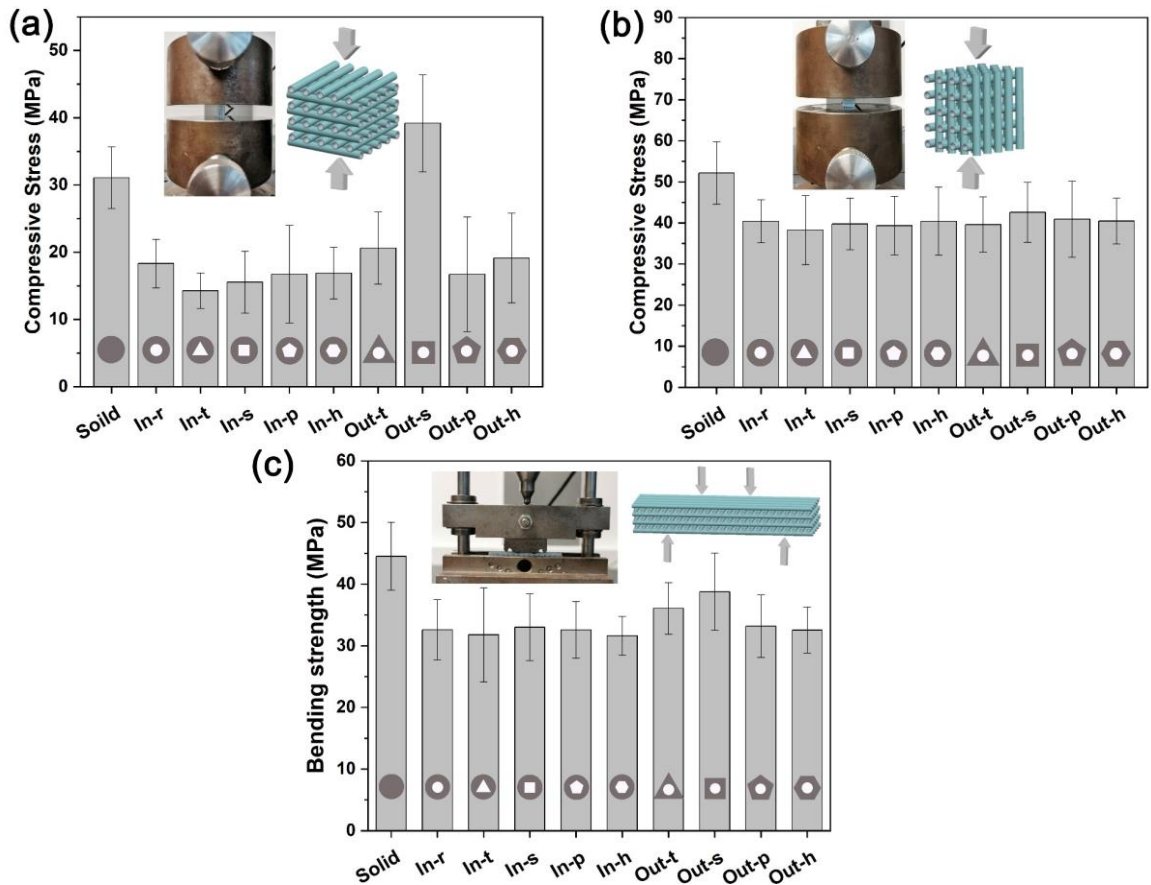

Fig. 5. Mechanical properties. Compressive strengths of the scaffolds, when compressed (a) in the transverse direction and (b) parallel to the direction of struts. (c) The four-point bend strengths of the scaffolds. (n=12).

In four-point bending tests of the Cu-DIO scaffolds, the solid strut scaffold had the highest bending strength of 44.5 ± 5.5 MPa (Fig. 5c). Hollow-strut scaffolds showed slightly lower bending strength values. It was seen that changing the outer shape of the hollow struts can produce an increase in bending strength, e.g., Out-t and Out-s have relatively high bending strengths of 36.1 ± 4.2 MPa and 38.8 ± 6.3 MPa, respectively. The fracture behavior is clearer in the representative load-displacement curves in Fig. S2c. All samples exhibited brittle failure, i.e. linear elastic response up to peak load followed by fracture.

To confirm the role of strut morphology in the reduction of stress concentration, FEA was conducted. Here, the simulations were accomplished in the compression and bending modes. Fig. 6a-f shows the contour of von Mises stresses of representative scaffolds under transverse and parallel compression (the von Mises stress contour plots of all scaffolds are shown in Fig. S3 and Fig. S4). Under transverse compression, all structures show stress concentration at interfaces between adjacent layers. Clearly, lower stress concentration occurs for square members as these interlayer contact areas are larger (Fig. 6c). Under constant compressive load, the unit cell of Out-s showed the lowest maximum deformation value of around 0.0011 mm, while these values were larger than 0.0031mm for other hollow strut structures (Fig. S5). Notably, the maximum deformation of the solid strut construct was also higher than that of Out-s, which was computed as 0.0018 mm. In parallel compression, FEA confirms that lower stress concentration occurs with compressive loads born by the parallel struts, while the stresses in horizontal struts are negligible. It is found that most of the constructs had a maximum deformation of around 0.0021 mm, except that the values of Solid and Out-s were calculated as a lower level, 0.0017 mm and 0.0020 mm, respectively (Fig. S6). FEA of four-point bending deformation for Solid, In-r, and Out-s scaffolds is shown in Fig. 6g-i (plots for four-point bending for all scaffolds are provided in Fig. S7). Under the applied loading Solid, Out-t and Out-s showed the greatest bending stiffness.

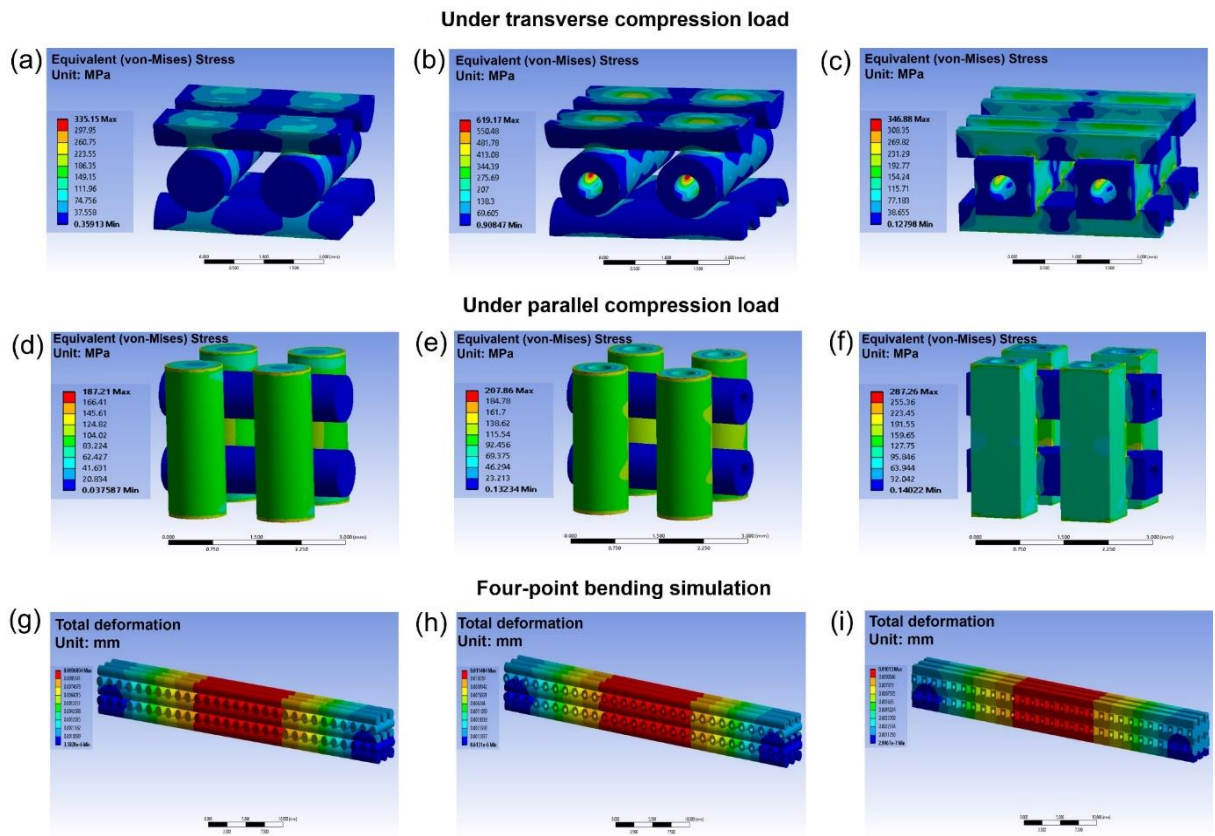

Fig. 6. Von Mises stress contour plot of representative scaffold under transverse compression load, (a) Solid, (b) In-r, (c) Out-s; and under parallel compression load, (d) Solid, (e) In-r, (f) Out-s. Total deformation contour plot of four-point bending simulation on the representative scaffold, (g) Solid, (h) In-r, and (i) Out-s.

### 3.4. *In vitro* cell experiments

The growth of bone depends, to a large extent, on the transportation of nutrients and metabolites in the vascular network. HUVECs are necessary for the establishment of new vascular networks, and osteoblasts play a key role in the growth and function of bone. Therefore, motivated by this, we studied how the presence of scaffolds affects the viability of HUVECs and Saos-2 cells. As shown in Fig. 7a, Cu-DIO extract had the same proliferation trend on HUVECS and Saos-2 cells after 1, 3, and 7 days of incubation. And the Cu-DIO significantly promoted cell growth compared to the control, demonstrating the good cytocompatibility of the material. Western blot was used to evaluate the expression level of CD31, which is considered as a marker for newly formed microvessels. The result of the western blot indicates that the Cu-DIO group exhibited a higher expression level than that of the control group at day 3 (Fig. 7b, c), revealing the greater effectiveness of Cu-DIO on angiogenesis.

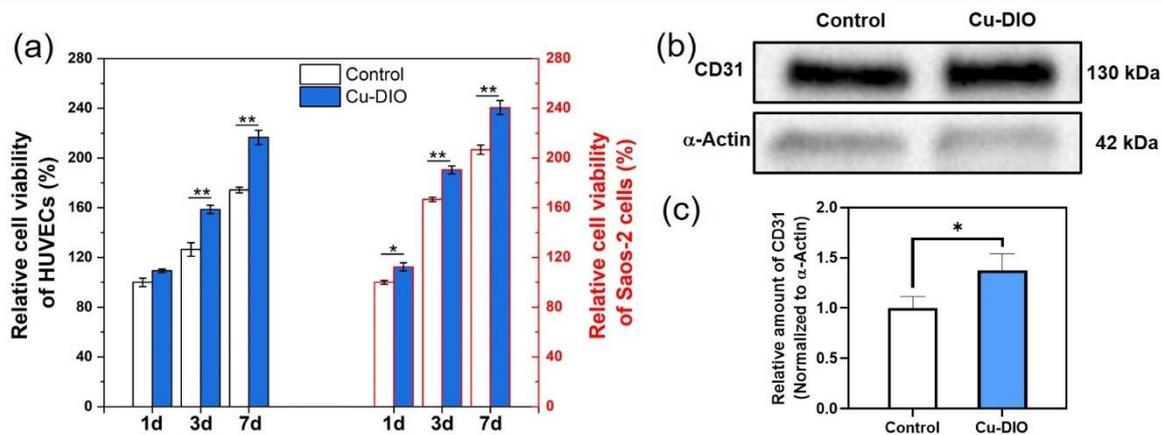

Fig. 7. *In vitro* cell experiment. (a) Cell viability after exposure to scaffold extracts for 1, 3, and 7 days. (n = 6, *P < 0.05 and **P < 0.01); (b) Representative protein expression bands of CD31 determined by western blot. (c) Western blot analysis quantification (n=3).

We further studied the growth of HUVECs and Saos-2 cells on different shaped hollow-strut scaffolds after live/dead staining. As shown in Fig. 8a, HUVECs adhered to both the outer and inner surfaces of hollow struts after 48 h of culture (as indicated by the arrow). In Fig. 8b the growth of Saos-2 cells on the scaffold was consistent with that of HUVECs, with cells adhering to the inner surface of the channel. Results showed that scaffolds with hollow channels provided more space for cell adhesion and proliferation compared to the solid scaffold, and no significant difference was found in the rest of the scaffold. Fig. 8c shows the Saos-2 cells inside the hollow struts after 7 days of culture. The results show that the culture medium can pass through the hollow channel to provide nutrients for the cells in the channel, so the cells have good attachment and viability inside the channels, which can be expected that the channel structure is able to transport nutrients and allow the growth of bone tissue into it. Although the mechanical properties are somewhat diminished, the inner channel surface area of In-t is nominally 28.6% greater than In-r, while in Out-t the outer strut surface area is also enhanced by around 28.6% relative to In-r, while having similar compressive performance. This illustrates how the selection of appropriate strut geometry allows a greater extent of bioactive interfaces without sacrificing mechanical performance.

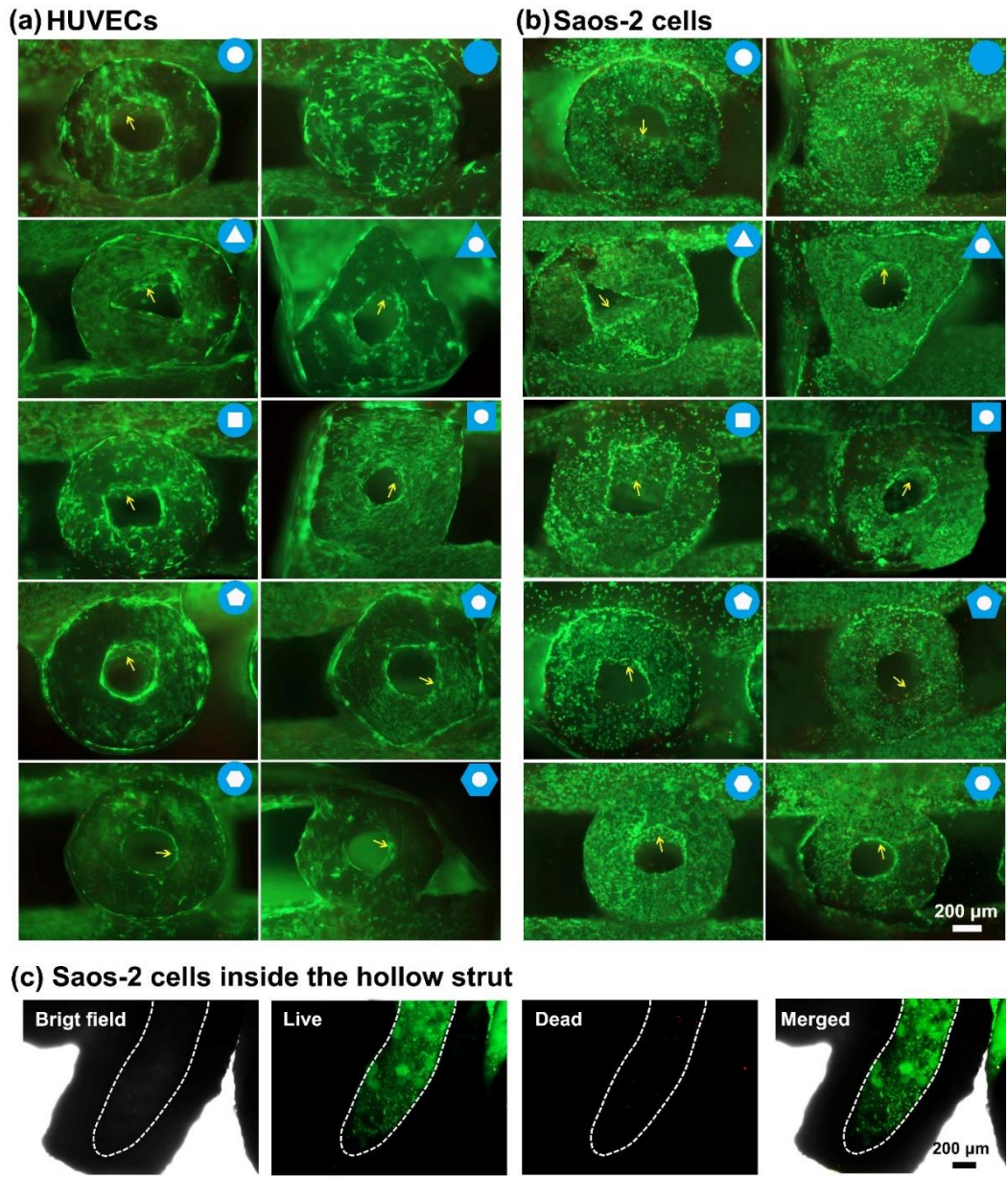

Fig. 8. Fluorescence images of cell viability assays of (a) HUVECs and (b) Saos-2 cells cultured on different shaped scaffolds for 48 h, and (c) Saos-2 cells inside hollow struts after 7-day culture, n=3.

## 3.5. Ceramic/polymer core-shell composite scaffolds

Scaffolds with hollow struts can be used as an intermediary for producing composite scaffolds with core-shell structured members. The hollow struts' channels can be penetrated with biodegradable polymers to produce a bioactive scaffold with outstandingly high toughness (Fig. 9a). A representative optical microscope image of the PCL-infiltrated In-r scaffold is shown in Fig. 9b. As shown in Fig. 9c, the composite scaffold showed continued load-bearing capacity even after a large strain, which is attributed to the contribution of PCL in the core. The proposed strategy provides enhanced strength without sacrificing the bioactivity of the scaffold in principle, which is an example of the application of bioceramic scaffolds with hollow structures.

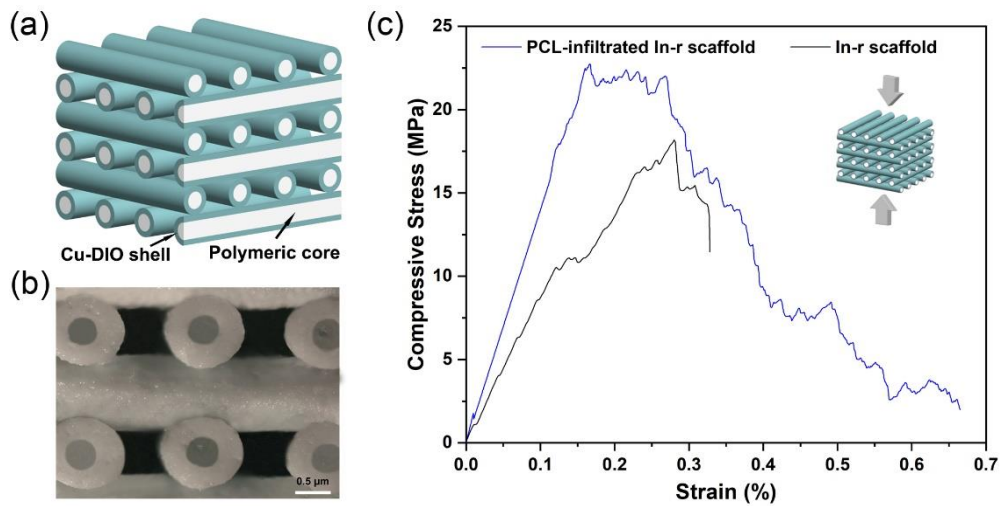

Fig. 9. Polymer-infiltrated scaffold production. (a) Schematic illustration of PCL-infiltrated scaffold, (b) optical microscopy image of cross-sections of the representative PCL-infiltrated Cu-DIO scaffold, and (c) representative strain-stress curve from compression test.

## 4. Discussion

The work presented here is motivated by prospects for improving patient outcomes in recovery from large-scale bone defects stemming from congenital malformations, tumors or infections. To effectively treat such bone defects, bone tissue engineering scaffolds require high levels of mechanical and bioactive performance. In this work, copper substituted diopside scaffolds with hollow members of varied geometries were simply printed by robocasting through specially designed extrusion nozzles, and their utility for bone repair was explored in terms of bioactivity and mechanical performance. The Cu-DIO used here for the fabrication of scaffolds has shown to be a promising material for bone tissue engineering scaffolds as it uniquely combines good fracture toughness with high levels of bioactivity. The weight loss of Cu-DIO after 8 weeks of immersion in Tris buffer was about 10.4%, indicating that the material is biodegradable [18]. The hollow channels and the surfaces therein provide also for greater biodegradability of scaffolds, and on the basis of this study, the adjustment of strut morphology can be envisaged as an approach to balancing scaffold degradation rate with mechanical support during bone tissue regeneration within the scaffold. We have shown that the morphology of hollow struts and their stacking directly affect the mechanical performance of scaffolds, which presents a promising pathway towards the enhancement of mechanical performance in engineered porous structures. Furthermore, inner channels of differing cross-sections further offer greater levels of bioactive surfaces available for the processes of osseointegration, which are necessary for the in-growth of human bone tissue in such scaffolds.

Known methods for the additive manufacturing of hollow strut based scaffolds are limited to nested needle type assemblies, which allow only tubular struts with circular channels [32]. Examination of images of scaffolds produced by such approaches reveals that deformation of the hollow struts is prevalent and channels are often not co-axial with struts [15]. In this work, in order to robocast scaffolds with continuously co-axial channels and minimal deformation, we produced integrated single-component extrusion nozzles by SLA 3D printing, a novel approach that allows the production of customizable strut morphologies. The designed nozzles used here have a conical geometry and a spatially fixed mandrel. Compared to coaxial extrusion in the prior art, generally involving two nested cylindrical needles of different sizes, those we created here allow a smoother extrusion of the slurry under lower pressure. The conical channel inside the presented nozzles is also beneficial to achieving a higher flow rate, maintaining a higher average velocity in slurry flow, and preventing clogging [33, 34]. To facilitate coaxiality of the mandrel, these are connected to the nozzle body via bridges having sharp upper surfaces. The acute angles on the bridge can further decrease the flow resistance during slurry extrusion. Ten different morphologies were successfully produced using the designed nozzles with continuous well-centered channels throughout, illustrating how this approach goes beyond the conventional design of scaffolds composed of round struts or struts with hollow round channels. The rigidity of the slurry used here is key to avoiding deformation pre- and post-sintering and obtaining well-controlled structures. The printing performance of these slurries can be further attributed to the additives used. A commonly used additive Pluronic F127 was used as a binder and support material due to its phase change properties, while sodium alginate provides further viscoelasticity due to its ability to cross-link with $Ca^{2+}$ ions in diopside to form a stable hydrogel. The addition of 1-octanol as a surfactant reduces bubble entrapment and improves the microstructure of the printed ceramic. Rheological analysis showed that the slurry had favorable shear thinning and shear recovery behavior similar to 40 wt% Pluronic F127, which can be considered as a standard for such 3D printing [27, 28]. The pseudoplastic shear-thinning alongside the aforementioned high values of low- to zero-shear viscosity of the slurries used here are key to the high fidelity of the hollow strut geometries fabricated here and avoiding slumping of struts and partial or complete closure of the channels, which have often been problematic in similar studies [31, 35].

Bone tissue engineering scaffolds require adequate mechanical strength during bone regeneration. The strength of human bone is a result of a strong dense outer bone (cortical bone) and a more porous inner material (cancellous bone). The compressive strengths of cortical and cancellous bone are reported as 100 - 230 MPa and 2 - 12 MPa, and the bending strengths are 50 - 150 MPa and 10 - 20 MPa, respectively [36]. We have shown here that hollow-strut scaffolds with continuous channels in a simple woodpile structure with high porosity of around 84% significantly exceed the mechanical performance of cancellous bone. Arguably, the hollow-strut diopside structures studied here are among the best-performing candidates for bioactive resorbable scaffolds in terms of specific strength. Although *in vivo* studies are not conducted here, the observed osseointegrative bioactivity of the substituted diopside materials combined with hollow channel structures is expected to facilitate the ossification of the inner regions of such scaffolds. Generally, the compressive strength of woodpile structure scaffolds is lower under transverse loading due to higher stress concentration. It can be seen from our study (mechanical testing and FEA) that, as expected, the outer shapes of the struts have much higher relevance to the scaffold's mechanical properties relative to their inner geometries. Interestingly, the Out-s scaffold exhibits significantly higher strength in compression (under transverse loading) and four-point bending. As confirmed by FEA, this is due to the larger contact areas between the quadrilateral strut layers, allowing for a uniform stress distribution that can effectively withstand the applied load. Achieving a simultaneous increase in porosity and compressive strength in hollow-strut scaffolds with square members is a valuable result that is most clearly manifested by the significant increase in specific strength. The specific strength reached in square strut scaffolds, averaged over both orientations, approaches 35.58 MPa·cm$^3$/g, which is higher compared to most porous titanium structures studied by Yánez et al. for human cancellous bone implant applications [37]. These results show that strut geometry can be used as a tool to impart greater mechanical strength in tissue engineering.

It is known that geometry significantly affects cellular responses and bone tissue regeneration [38-41]. Entezari et al. [11] concluded that the pore of the scaffold should be greater than 390 μm with a maximum of 590 μm to improve bone tissue formation, and hollow strut scaffolds exhibit particular benefits for tissue regeneration [14]. Hollow channels not only boosted HUVEC capillary-like tube development, but also facilitated host vascular infiltration deep into the scaffolds *in vivo*, according to Zhang et al. [42]. In our work, it was confirmed that cells adhere to both the outer and inner surfaces of struts. The presence of channels within the struts increases the extent of available surfaces in the scaffolds examined here by 40.1 - 51.4 %, providing more surface for cell adhesion and enhancing osseointegration. In particular, the Cu-DIO material used here, as indicated by XTT and western blot analysis (Fig. 7), has good levels of biocompatibility and angiogenic bioactivity. The inflammatory response can be influenced by the scaffold shape, according to Almedia et al., with larger pores resulting in higher quantities of pro-inflammatory cytokines secreted [43]. The role of pore geometry was also discussed by Xu et al., who observed that scaffolds with a parallelogram pore geometry showed the strongest alkaline phosphate activity compared to square and triangular pore shapes [44]. However, the exact mechanism that promotes or inhibits cell adhesion, proliferation, and differentiation in scaffolds with various pore shapes is yet unknown. Ascertaining how transport through channels and cell attachment on channel walls combine to influence the *in vivo* performance of additively manufactured tissue-engineering scaffolds with hollow struts of varying cross-section remains ripe for further exploration.

## 5. Conclusions

We report here a new design of coaxial extrusion nozzles that allows the robocasting of hollow-strut scaffolds with varied member cross-sections. We investigated diopside based scaffolds produced using this technique towards applications in bone tissue engineering. The hollow-strut copper substituted diopside (Cu-DIO) scaffolds here have high porosity and mechanical strength superior to cancellous bone. Due to lower stress concentrations in the woodpile structure, scaffolds consisting of square cross-sectional struts with circular inner channels (Out-s) show the highest compressive strength and impressive bending strength compared to other scaffolds both hollow and solid. The biocompatibility and angiogenic activity of the Cu-DIO material combined with the unique hollow channels of the scaffold greatly improve cell attachment and proliferation that could further promote the formation of new bone within the hollow struts of the scaffold *in vivo*. Overall, this study provides a practical strategy for the additive manufacturing of hollow channel scaffolds, which have a broad range of potential applications not only in tissue engineering but also in catalysis, environmental materials as well as precursors toward core-shell composite structures.

## Declaration of competing interest

The authors declare that they have no known competing financial interests or personal relationships that could have appeared to influence the work reported in this paper.